\documentclass[twocolumn,amsmath,floats,showpacs,nofootinbib]{revtex4}
\usepackage[usenames]{color}
\usepackage{graphicx}
\usepackage{epstopdf}
\usepackage{textcomp}
\usepackage{hyperref}
\usepackage{multirow}
\usepackage{tikz}
\usepackage{rotating}

\hypersetup{colorlinks=true}

\begin{document}

\title{Excess conduction of YBaCuO point contacts between 100 and 200 K}

\author{L. F. Rybal'chenko, I. K. Yanson, N. L. Bobrov, V. V. Fisun, M. A. Obolenskii*, A. V. Bondarenko*, Yu. D. Tret'yakov**, A. R. Kaul'**, and I. E. Graboi**}
\affiliation{Physicotechnical Institute of Low Temperatures, Academy of Sciences of the Ukrainian SSR, Kharkov,\\
A. M. Gorky State University, Kharkov*,\\
and M. V. Lomonosov State University, Moscow**\\
Email address: bobrov@ilt.kharkov.ua}
\published {(\href{http://fntr.ilt.kharkov.ua/fnt/pdf/16/16-1/f16-0058r.pdf}{Fiz. Nizk. Temp.}, \textbf{16}, 58 (1990)); (Sov. J. Low Temp. Phys., \textbf{16}, 30 (1990)}
\date{\today}

\begin{abstract}$YBaCuO-Ag$ pressure point contacts with direct conduction are investigated. The excess (relative to the normal state) conductivity mainly caused by fluctuational pairing of electrons above $T_c$ is measured in the temperature interval 100-200~$K$. The superconductivity above 120~$K$ is found to be of the two-dimensional type. The obtained preliminary results indicate the presence of small amount of an unknown phase with $T'_c\gtrsim 200~K$ in $YBaCuO$.


\end{abstract}

\maketitle

Fluctuational electron pairing above the superconducting transition temperature $T_c$ is known to lead to an additional contribution to electrical conductivity, whose fluctuational dependence on temperature can be used to determine the geometrical dimensionality of the object under investigation or of an individual subsystem responsible for superconductivity \cite{1}. In view of the small charge carrier concentration in new high-temperature superconductors (HTS), the fluctuational correction $\sigma'=[\sigma(T)-\sigma_N]$ to conductivity ($\sigma(T)$ is the conductivity being measured and $\sigma_N$ the conductivity corresponding to the normal state) is considerably larger than for traditional superconductors, which causes, for example, noticeable "rounding" effects in the resistance dependences $R(T)$ above $T_c$.

The measurements of fluctuational corrections to the conductivity $\sigma(T)$ and to the magnetic susceptibility $\chi(T)$ made in a number of works \cite{2,3,4} not only for ceramic, but also monocrystalline $YBa_2Cu_3O_{7-\delta}$ samples did not provide an unambiguous information about the dimensionality of the superconducting subsystem. For example, the authors of Ref. \cite{2} reported about three-dimensional fluctuations observed in this compound at $\tau>0.1
(\tau=T/T_c-1)$, while in Ref. \cite{3} these fluctuations are regarded as two-dimensional. The reason behind such contradictory conclusions apparently
lies in insufficient homogeneity of investigated samples including monocrystalline ones. Due to the presence of a large number of weak bonds between individual structural elements (granules, grains, and subcrystallites) in a sample, these bonds make a considerable contribution to the signal being recorded, which considerably complicates the determination of the type of superconducting fluctuations within stoichiometry regions.

In order to minimize the effect caused by sample heterogeneity, we measured the excess conductivity in $YBa_2Cu_3O_{7-\delta}$ by using pressure points contacts (PC) of the S-N type (N=$Ag$) with the metal type conductivity, thus restricting the region under investigation to submicron size. It turned out that in many cases superconducting effects are observed up to 220-230~$K$, and in the temperature interval from about 120 to 180~$K$ the excess conductivity obeys the law typical of two-dimensional systems. The noticeable effect of the magnetic field on excess conductivity at $T>2T_c$ suggests that there exist small amounts of unidentified superconducting phase with $T_c\gtrsim 200~K$.

\begin{figure}[]
\includegraphics[width=8cm,angle=0]{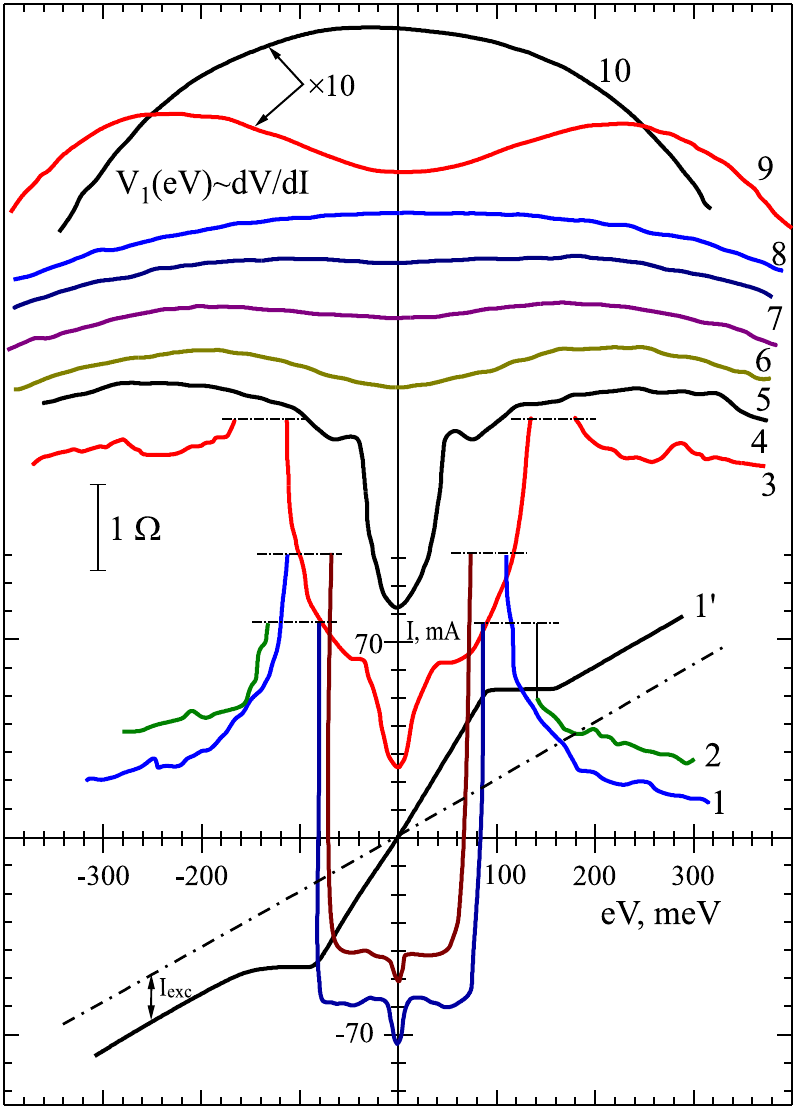}
\caption[]{Dependence of the differential resistance on voltage (curves 1-10) and IVC (curve 1') for the $YBaCuO-Ag$ point contact at different temperatures $T, K$: 20(1,1'), 40(2), 80(3), 90(4), 100(5), 125(6), 155(7), 220(8), 180(9) and 230(10). The dot-and-dash line is the expected IVC in the N-state.}
\label{Fig1}
\end{figure}

Point contacts were prepared by pressing a sharpened N ($Ag$) electrode to a freshly fractured surface of the $YBa_2Cu_3O_{7-\delta}$ sample. In some cases, the sample was in the form of a ceramic pellet prepared according to cryochemical technique \cite{5}, and in other cases it was in the form of a single crystal of size $\sim 1\times 1\times 0.2~mm^3$ grown by the solution-melt technique. In both cases, the experimental results were close. The design
of the device used for PC studies was such that the point of contact could be moved over the fractured surface during one measuring cycle with the help of a mechanical drive fixed to the cryostat from outside. As was pointed out earlier \cite{6}, such a technique allowed us to determine individual regions on fresh fractures where the critical parameters attain high values ($T_c=90-95~K$). The excess conductivity above 100~$K$ was observed in the present work just in such regions. It should be emphasized that the excess conductivity is manifested only when there is a considerable excess current $I_{exc}$ on the current-voltage characteristics (IVC) of the corresponding contacts (Fig. \ref{Fig1}, curve 1'). This is an indication of the predominant filling of the PC region with a superconducting phase and of the absence of barrier layers with a low transparency at the contact between electrodes.

The excess conductivity of point contacts caused by superconductivity was manifested in the form of a broad symmetric minimum near $V=0$ on the voltage dependences of differential resistance $R_D(V)\propto dV/dI(V)$ above 100~$K$ (see Fig. \ref{Fig1}). The differential resistance of a point contact was determined from the first harmonic of a weak modulating signal $V_1(V)\propto R_D(V)$ of frequency 487~Hz in a bridge circuit with a current source. The relative error of measurements was not higher than $10^{-4}$. The error in the measurement of temperature above 100~$K$ was $\sim 0.5~K$. It should be noted that when the temperature decreases below $T_c$, the intensity of this minimum increases considerably due to the Andreev electron reflection initiated by the emergence of an energy gap in the superconducting electrode. In this case, the principal minimum acquires an additional structure in the form of weak gap minima at $\pm eV=\Delta\simeq 40~meV$ and a narrow minimum at $V=0$, whose nature will not be discussed here.

The typical values of the differential resistance for the investigated contacts at $eV\gg\Delta$ were $R_N\sim 10~\Omega$. Using the Maxwell formula for estimating the diameter $d_{pc}$ of the point contact constriction and neglecting the contribution of the silver bank to $R_N$, we obtain $d_{pc}=\rho_{YBaCuO}/ 2R_N\sim 500~\text{\AA}$, considering that the mean values of the resistivity $\rho_{YBaCuO}$ of the ceramic under
consideration are close to $10^{-4}~\Omega \cdot cm$. The PC characteristics are mainly determined by the contact constriction (which is small for our samples), and hence we can assume that the effect of intercrystallite weak coupling on the excess conductivity is reduced to a minimum in our experiments.

\begin{figure}[]
\includegraphics[width=8cm,angle=0]{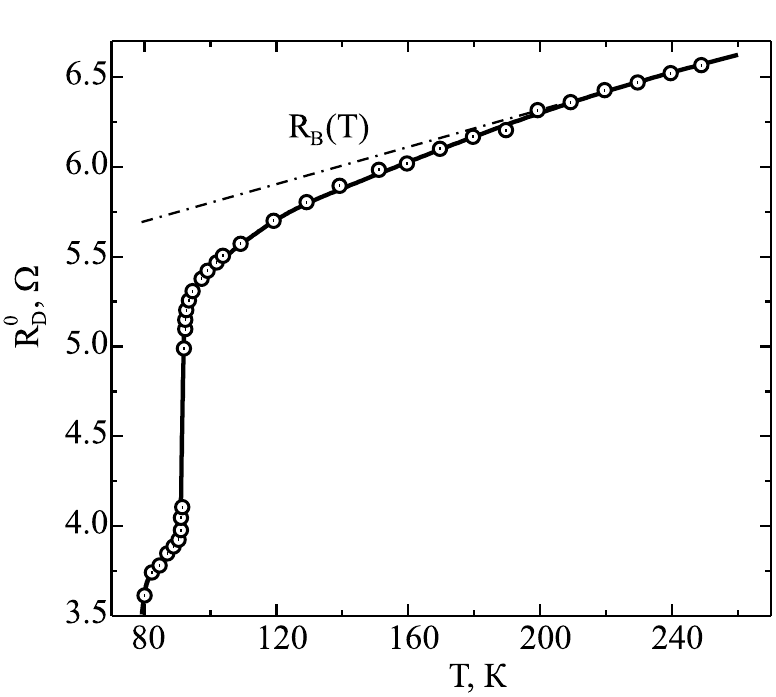}
\caption[]{Temperature dependence of the differential resistance $R_D^{(0)}$ for $YBaCuO-Ag$ point contact at $V=0$ (the dot-and-dash line is the approximation of $R_D^{(0)}(T)$ at $T>200~K$, i.e., $R_B(T)$.}
\label{Fig2}
\end{figure}

The value of the excess conductivity ${\sigma }'(T)$ approximately corresponds to the depth of the central minimum (near $V=0$) on the $dV/dI$ curves (see Fig. \ref{Fig1}). The exact values of ${\sigma }'(T)$ were obtained by using another method. At first,
the $R_D^{(0)}(T)$ dependence (the differential resistance of PC at $V=0$) was registered in a wide temperature range (Fig. \ref{Fig2}). Then this dependence was approximated in the high-temperature region by the straight line $R_B(T)$ which is supposed to reflect the background change in the resistance of the bulk conductor. The absolute value of the excess conductivity at each temperature value was determined by using the formula
\[{\Sigma }'(T)=\frac{1}{R_{D}^{(0)}(T)}-\frac{1}{{{R}_{B}}(T)}\]

\begin{figure}[]
\includegraphics[width=8cm,angle=0]{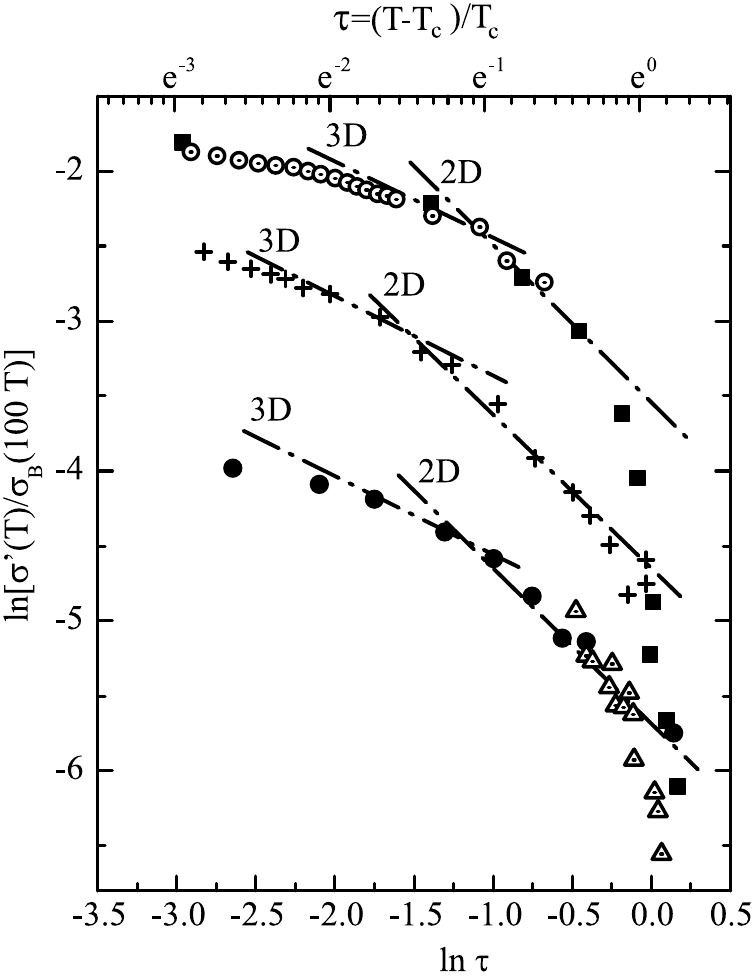}
\caption[]{The normalized excess conductivity ${\sigma }'(T)$ at $V=0$ for several $YBaCuO-Ag$ point contacts as a function of reduced temperature in logarithmic coordinates (the dot-and-dash lines show the expected slope of these dependences for three- and two-dimensional fluctuations; different symbols correspond to different contacts).}
\label{Fig3}
\end{figure}

According to Ref. \cite{1}, the fluctuational conductivity in traditional superconductors is $\sigma_{AL}(T)=A\tau^{-K}$, where $K$=1/2; 1; 3/2; and 2, respectively, for three-, two-, one-, and zero- dimensional objects respectively, $\tau=T/T_c-1$, and $A$ is a temperature-independent coefficient. It was

shown by Maki \cite{7} and Thompson \cite{8} that the formation of fluctuational pairs in the pure limit is also responsible for a considerable increase in the conductivity of unpaired electrons. However, the Maki-Thompson contribution can be disregarded in materials with a short electron mean free path (like HTS). Figure \ref{Fig3} shows in logarithmic coordinates the results of measurements of the excess conductivity $\sigma'(T)$ of several point contacts, which is normalized to $\sigma_B$ (at 100~$K$), as a function of the reduced temperature $\tau\ (T_c=90-95~K$ was determined from the midpoint of a sharp change
$R_D^{(0)}(t)$ as, for example, in Fig. \ref{Fig2}). It was taken into account that ${\sigma}'(T)/\sigma_B(100~K)={\Sigma}'(T)/\Sigma_B(100~K)$, where $\Sigma_B(100~K)=1/R_B (100~K)$.

In spite of the considerable spread in the values of the ratio ${\sigma}'(T)/\sigma_B(100~K)$ for different contacts, which can be attributed to variations
of the chemical composition of the S-electrode in the region of PC constriction, the ${\sigma}'(T)$ dependence above 120~$K$ approximately obeys a power law with the exponent $k=1$. As the temperature approaches $T\sim2T_c$, this law is violated, but it is impossible to determine the behavior of ${\sigma}'(T)$ in this temperature region. With lowering temperature, the 2D-dependence is replaced by the 3D-mode which is manifested only in a narrow temperature interval (10-15~$K$). With a further decrease in temperature, the power law is modified, in particular, due to an increase in the contribution from strong (critical) fluctuations in the immediate vicinity of $T_c$, which are not analyzed in this work.

Thus, an analysis of Fig. \ref{Fig3} leads to the conclusion that weak superconducting fluctuations in $YBa_2Cu_30_{7-\delta}$ far from $T_c$ are two-dimensional
 as should be expected in view of the anisotropic structure of this substance. In the Aslamazov-Larkin formula for 2D-systems, $A=e^2/16~\hbar d$. Having
determined the value of $A/\sigma_B(100~K)$ from Fig. \ref{Fig3} and assuming that $d=12~\text{\AA}$, we can calculate the resistivity of the PC region which is found to be $\rho_B(100~K)\sim (0.3-3)\cdot 10^{-4}~\Omega\cdot cm$. This value is in reasonable agreement with the data on bulk material. Strictly speaking, an analysis of experimental results for layered superconducting compounds like $YBaCuO$ should be carried out by using the Lawrence-Doniach model \cite{9}. However, far away from $T_c(T\gg T_c)$ this model leads to an expression which does not differ much from ${\sigma}'_{AL}(T)$ , which confirms the correctness of our conclusions.

\begin{figure}[]
\includegraphics[width=8cm,angle=0]{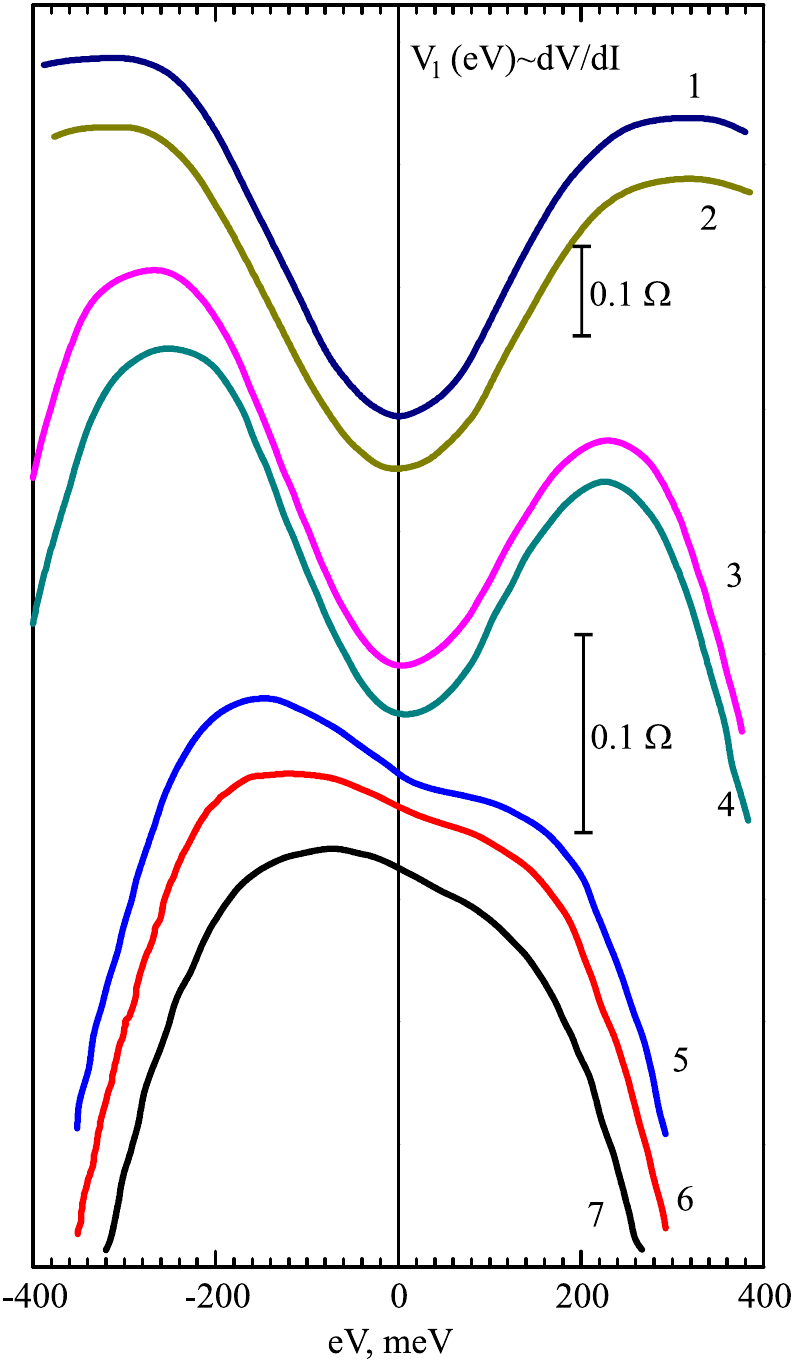}
\caption[]{The effect of the magnetic field on the $dV/dI$ characteristics of $YBaCuO-Ag$ point contact: $H=0$, $T=150~K$, $R_D^{(0)}=6.8~\Omega$ (1);
 $H=43~kOe$, $T=150~K$ (2); $H=0$, $T=180~K$ (3); $H=49~kOe$, $T=180~K$ (4); $H=0$, $T=220~K$ (5); $H=47~kOe$, $T=220~K$ (6); and $H=0$, $T=230~K$, $R_D^{(0)}= 7.3~\Omega$(7).}
\label{Fig4}
\end{figure}

We also analyzed the effect of a magnetic field on the excess conductivity of the investigated point contacts (Fig. \ref{Fig4}). A comparison of the $dV/dI$ characteristics in different fields and at different temperatures shows that the maximum field-induced relative changes in the depth of the central minimum (which directly correlates with the excess conductivity) occur far from $T_c$, i.e. at $T\sim 220K~$, which does not agree with fluctuational theory of paraconductivity developed by Hikami and Larkin for layered superconductors \cite{10}. As a matter of fact, according to Ref. \cite{10} relatively weak fields (of the order of several tens kilooersteds) cannot
considerably affect $\sigma'(T)$ away from $T_c$, and this effect becomes weaker with increasing temperature. On the contrary, a sharp increase in the magnetoresistance can take place only in the immediate proximity of $T_c$. Indeed, the experiments carried out by Freitas et al. \cite{2} revealed that the positive magnetoresistance is $\Delta\rho/\rho\sim 0.1-0.2\%$ in the field 40~$kOe$ at $\tau=0.2$. Consequently, it cannot be ruled out that a small amount of an unknown phase with $T_c\gtrsim 220~K$, which makes a certain contribution to the paraconductivity observed above 100~$K$, gets in the region of constriction in the $YBaCuO$-based point contacts studied in the present work.

\begin{figure}[]
\includegraphics[width=8cm,angle=0]{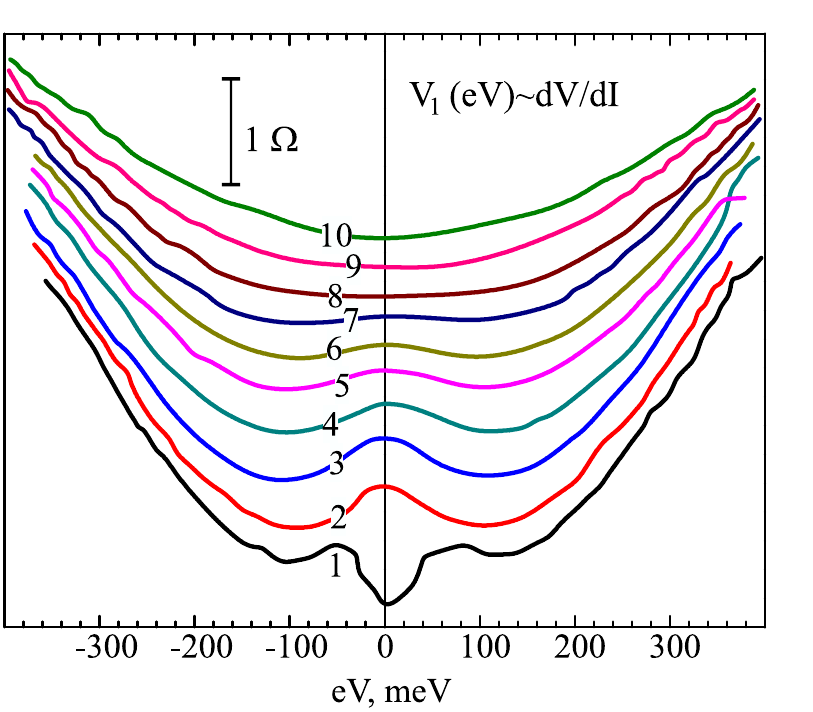}
\caption[]{Gap-like minima on the $dV/dI$ dependences for $YBaCuO-Ag$ point contact at different values of $T,\ K$: 90(1), 96(2), 104(3), 120(4), 130(5), 142(6), 150(7), 160(8), 166(9) and 170(10) (the contact resistance $R_D^{(0)}(96~K)\simeq 4 ~\Omega$ at $T=96~K$ and $V=0$).}
\label{Fig5}
\end{figure}

In connection with the last assumption, it is also interesting to analyze the $YBaCuO$ characteristics of PC, obtained on a monocrystalline $dV/dI(V)$ sample (Fig. \ref{Fig5}). As in the prevous cases, a considerable excess conduction is observed, but this case is distinguished by the emergence of broad gas-like minima on the $dV/dI$ dependences near $\pm eV\simeq 100~meV$. The position of these minima on the energy axis does not depend on temperature. We can assume that the constriction region of the point contact presented in Fig. \ref{Fig5} contains a larger amount of the unknown phase mentioned above or that this phase is closer to the center of the contact than in the cases considered earlier. Then the values of energy corresponding to the minima on the $dV/dI$ dependences determine the value of the energy gap in the unknown phase. The absence of the temperature dependence of the positions of these minima on the energy axis can be related to the specific nature of this phase as, for example, in the Kulik theory \cite{11} where the situation when the gap of the pairs does not depend on temperature is analyzed.

Thus, the indications of the presence of small amounts of a new unidentified phase with $T'_c\gtrsim 200~K$ were obtained in the present work along with the observation of the quasi-two-dimensional nature of fluctuational superconductivity in $YBaCuO$.

\begin{table}[h!]
  \centering
\begin{flushleft}
\textbf{Paraconductivity of YBaCuO point contacts above 100 K}\\

\emph{\textbf{\textbf{I. K. Yanson, L. F. Rybalchenko, N. L. Bobrov, and V. V. Fisun}\\
Institute for Low Temperature Physics and Engineering, Kharkov, \href{http://fntr.ilt.kharkov.ua/fnt/pdf/16/16-5/f16-0661ea.pdf}{Fiz. Nizk. Temp.}  16, 661 (May 1990)\\
1 Soviet - West-German Symposium on NTSC\\
}}
Due to the low charge-carrier concentration in the new high-$T_c$ superconductors, the fluctuating pairing of electrons\\ at temperatures above $T_c$ is much stronger than in conventional superconductors. In this study we observed excess conductivity of point contacts between $YBa_2Cu_3O_{7-y}$ and $Ag$ at temperature above $T_c$. Contacts having a resistance $R_N15<\Omega$ were made by the shear method inside the cryostat. At $T<T_c$ ($T_c=92-94~K$) they were characterized by a large excess current, which indicates that the constriction region at the S-electrode side consisted mainly of the superconducting 1-2-3 phase and there were no apppreciable barrier layers at the electrode interface. The excess conductivity of such contacts appeared as a wide minimum on the $dV/dI(V)$ dependences at temperatures above $T_c$ up to 200-220~$K$. In this temperature range we calculated the excess conductivity at $V=0$ and $\sigma'=1/R_{T}-1/R_N$, where $R_{T}$ is the contact resistance at a given temperature and $R_N$ is the normal state resistance. Plotting these results in logarithmic coordinates $\sigma'/\sigma_N-\tau(\tau=T/T_c-1)$, we find that in the interval 120-180~$K$ the excess conductivity can, with good accuracy, be respresented by the power law $\sigma'/\sigma\propto \tau^{-1}$ characteristic of the two-dimensional systems. The noticeable effect of the magnetic field on the excess conductivity at $T>2T_c$ enables one to postulate the presence of small quantities of an unknown superconductive phase with $T_c>200~K$.
  \end{flushleft}
\end{table}

\end{document}